# Feature Selection of Post-Graduation Income of College Students in the United States

Ewan Wright[1], Qiang Hao[2], Khaled Rasheed[3], & Yan Liu[4]

[1] University of Hong Kong, Hong Kong SAR
[2] Western Washington University, Bellingham, WA, United States
[3] University of Georgia, Athens, GA, United States
[4] University of British Columbia, Canada

**Abstract.** This study investigated the most important attributes of the 6-year post-graduation income of college graduates who used financial aid during their time at college in the United States. The latest data released by the United States Department of Education was used. Specifically, 1,429 cohorts of graduates from three years (2001, 2003, and 2005) were included in the data analysis. Three attribute selection methods, including filter methods, forward selection, and Genetic Algorithm, were applied to the attribute selection from 30 relevant attributes. Five groups of machine learning algorithms were applied to the dataset for classification using the best selected attribute subsets. Based on our findings, we discuss the role of neighborhood professional degree attainment, parental income, SAT scores, and family college education in post-graduation incomes and the implications for social stratification.

## 1    Introduction

Higher education is an excellent "investment" that should be encouraged by families, schools, communities, and policy makers. The returns of a college degree vis-à-vis a high school diploma has expanded considerably in recent decades. Autor [1] demonstrates that this "graduate premium" doubled in real terms between 1979 and 2012. The gap in earnings between the median college educated worker and the median high-school educated worker increased from $17,411 to $34,969 for men, while also increasing from $12,887 to $23,280 for women. This reflects a rising demand for skills in the U.S. labor market. On the one hand, technological advancements are viewed as complementing the productivity of highly-educated workers while simultaneously automating routine jobs of lowly-educated workers [2]. On the other hand, the globalization of production has resulted in the relocation of jobs in certain "blue collar" industries exposed to import competition [3]. Indeed, research by Chetty et al. [4] underscores the role of higher education as a key pathway to intergenerational social mobility in the U.S. Further, Hout [5] contends that higher education "makes life better" through a host of social benefits in community relations, health, family stability, and social connections.

The proportion of 18 to 24-year olds enrolled at degree granting higher education institutions in the United State increased steadily from 25.7 percent in 1970 to 32.0 percent by 1990 and then to 40.5 percent in 2015, although enrolment growth at four-year institutions has stagnated in recent years [6]. A corresponding trend has been a rise in the costs of higher education [7]. Wolff et al. [8] estimate that higher education tuition fees have risen by 250 percent since the early 1980s when measured in dollars of constant purchasing power. The increases in cost have often not been met by a corresponding increase in grants or other forms of aid, especially as state funding for higher education has come under pressure [9]. To fill this funding gap, students have become more dependent on student loans to finance higher education studies. Brown et al. [10] demonstrate that the proportion of 25-year olds with student debt grew from 27 percent in 2004 to 43 percent in 2012 (p. 7), while during the same period the average debt being held grew by 70 percent to $25,000 (p. 5).

Moreover, as higher education participation has expanded, college graduates have become an increasing heterogeneous population with increasingly disparate labor market outcomes [11, 12]. While some graduates are highly successful, others face challenges to gainful employment. Data from the Federal Reserve Bank of New York [13] puts unemployment among college graduates aged between 22 and 27 at 3.9 percent. More significantly, the same dataset shows that 43.5 percent of these recent graduates were under-employed or employed in a job that "typically does not require a college degree". This latter point is reflected in graduate earnings with the dataset also illustrating that 13.3 of recent graduates are employed in "low-wage jobs" that tend to pay below $25,000 per annum. Relatedly, Haughwout et al. [14] identified that half of college graduates from the 2009 cohort who took student loans have defaulted, gone delinquent, or made no progress in paying their debt.

In this context, it is important to understand who benefits most and least in the labor market from higher education. Research has established that major and institutional selectivity are important factors in post-graduation incomes. Hoekstra [15] identified a 24 percent earnings premium for graduating from a flagship state university, while Witteveen and Attewell [16] found that earnings were 21 percent lower for graduates of the least selective colleges relative to the most selective colleges. Further studies find large premiums for selective colleges for graduates from less-educated families and among Black and Hispanic graduates [17] and low-income families [4]. There is also evidence that graduates of majors associated with science, technology, engineering, and math (STEM) enjoy higher salaries and lower incidents of unemployment, compared to graduates from non-technical majors [18, 19]. For example, Kim et al. [19] estimate that lifetime median earnings gains for STEM degree holders ($800,000) are more than double the relative gains for a social science degree ($374,000).

Building on literature, this study explored the most important attributes of 6-year post-graduation income of college graduates who used student aid from the U.S. Department of Education, and to what extent of accuracy the selected attributes can be used to classify post-graduation income. The research questions are:

(1) What are the most important attributes of post-graduation income of college students who graduate with debt repayment obligations?

(2) To what extent can the selected attributes classify post-graduation income of college students who graduate with debt repayment obligations?

## 2 Research Design

### 2.1 Data Collection

The data for this study was the latest dataset – released in October 2015 – by College Scorecard under the U.S. Department of Education [20]. This dataset only covered students who used financial aid during their college study period. Each row in the data stands for a student cohort admitted to a certain university. The data ranged from 1996 to 2013, but the 6-year post-graduation income data are only available for the years 1997, 1999, 2001, 2003 and 2005. The response variable in the present study is the mean value of the 6-year post-graduation income of a student cohort. Attributes were filtered based on domain knowledge. Those deemed less relevant were excluded, such as latitude of the institution and percent of students who passed away within 6 years after graduation.

30 potential attributes (see Appendix A) under five groups were included in this project. The groups are: (1) School, (2) Admission, (3) Cost, (4) Student Cohort, and (5) Socioeconomic Status of Students. Some attributes in certain groups are not available before 2000, such as admission rate in the Admission Group. Thus, only three years of data, including 2001, 2003, and 2005 were used. 1,429 cohorts were included for the data analysis. The response variable, mean income value of each cohort, was discretized into four classes based on the American Individual Income Distribution; including Very low (0 to 25,000), Low (25,000 to 37,500), Middle (37,500 to 50,000), and High (Above 50,000) [21].

### 2.2 Data Analysis

Two steps of preprocessing were applied to the collected data before the analysis: (1) *Standardization*: Standardization, transforming raw scores to z-scores, was applied to all the numerical attributes. There were 28 numerical attributes in total; (2) *One-hot encoding*: One-hot encoding techniques were applied to all the nominal attributes. There were 2 nominal attributes.

The data analysis had two phases. Phase One was attribute selection. Three attribute selection methods were applied and compared, including filter methods, stepwise wrapper methods, and naturally inspired algorithms. The filter methods applied in this study included five algorithms: (1) OneR algorithm, (2) Relief-based selection, (3) Chi-square selection, (4) Gain-ratio-based selection, and (5) Information-gain-based selection.

Both stepwise wrapper methods and naturally inspired algorithms need to have an evaluation function to work. Logistic regression was chosen as the evaluation function of both for stability and efficiency. The stepwise wrapper methods include forward and backward selection. Forward selection starts with no attributes in the model, and tests the addition of each attribute using certain comparison criteria. Backward selection starts with all candidate attributes, and tests deletion of each attribute using certain criteria. Only forward selection was used in this study.

The naturally inspired algorithm implemented was the Genetic Algorithm. Genetic Algorithm is a computational algorithm with origins in the field of biology. The tools that Genetic Algorithm uses have marks of genetic systems, including generation selection, crossover, and mutation [22]. We implemented the simple form of Genetic Algorithm described by Goldberg [23].

Phase Two was the exploration of the extent to which selected attributes classified post-graduation income, and how they perform compared to the whole attribute set. 10 machine learning algorithms in five groups were applied to the dataset. The five groups of algorithms included:

(1) Bayes-based algorithms [*Naive Bayes Update, Bayes Net*];

(2) Function-based algorithms [*Logistic Regression, Support Vector Machine, Multilayer Perceptron*];

(3) Instance-based algorithms [*Distance-weighted K-Nearest Neighbor*];

(4) Tree-based algorithms [*J48 decision tree, Multiclass Alternating Decision Tree*]; and

(5) Rule-based algorithms [*OneR, JRIP*].

Weighted average F1-score was chosen as the primary evaluation criterion, because there exists an imbalance in the four income classes. A classifier that primarily guesses based on the majority class would achieve a small advantage in accuracy, but would perform worse in terms of the F1-score. Also, classification accuracy rate was used as the secondary evaluation criterion. Ten-fold cross validation was used for the estimation of both F1-score and accuracy rate.

## 3    Results

### 3.1    Attribute Selection

**Attribute Selection using Filter Methods.** Five filter methods were applied to the attribute selection: (1) OneR algorithm, (2) Relief-based selection, (3) Chi-square selection, (4) Gain-ratio-based selection, and (5) Information-gain-based selection. The 10-fold cross validation scheme was implemented in Weka [24]. In contrast to the cross-validation in prediction or classification, no training or testing is involved in the cross-validation scheme of attribute selection. Under such a scheme, the dataset was randomly sectioned into 10 folds, and only 9 folds were used for subset attribute selection in each round. There were 10 rounds in total. The 10 selection results were summarized afterwards. The attributes selected by at least three out of the five methods (60%) were selected, yielding 14 selected attributes. The arithmetic mean of each attribute's ordinal ranking across all selection methods was also calculated, to enable measuring of attribute usefulness. For each single-attribute evaluator, the output of Weka showed the average merit and average rank of each attribute over the 10 folds (see Table 1).

**Table 1.** Selected Attributes Subset using Filter Methods.

| Attributes | Votes*/ Average Rank* | Attributes | Votes*/ Average Rank* |
|---|---|---|---|
| % of Population from Students' Zip Codes over 25 with a Professional Degree | 5/ 2.88 | Admission Rate | 5 / 12.42 |
| Average Faculty Salary | 5 / 3.50 | Instructional Expenditure per Student | 4 / 7.25 |
| Average SAT Score | 5 / 5.22 | % of Students Whose Parents have Post-High School Degree | 4 / 9.23 |
| Degree Completion Rate | 5 / 6.10 | Out-of-State Tuition Fee | 4 / 10.18 |
| % of Asian Students | 5 / 7.22 | % of Students whose Parents were 1st Generation College Student | 4 / 10.33 |
| % of Students Whose Parents Have a High School Degree | 5 / 8.58 | % of 1st Gen. College Students | 4 / 10.63 |
| In-State Tuition Fee | 5 / 10.88 | % of Students whose Family Income classified Very High | 4 / 11.30 |

*Votes Column: The number of filter methods that selected the corresponding attributes; Average Rank Column: The averaged rank values among the filter methods that selected the corresponding attributes.*

**Attribute Selection using Forward Selection.** Same as the implementation of filter methods, 10-fold cross validation scheme in Weka was used for more stable estimates. Attributes selected by at least six out of ten folds (60%) were selected, yielding 9 selected attributes in total. The selected attributes are presented in Table 2.

**Table 2.** Selected Attribute Subset using Forward Selection

| Attributes | Votes* | Attributes | Votes* |
|---|---|---|---|
| Predominant Degree Type | 90% | % of Students whose Parents were 1st Generation College Student | 60% |
| Ratio between Part-time and Full-time Students | 100% | % of the Population from Students' Zip Codes over 25 with a Professional Degree | 100% |
| Degree Completion Rate | 100% | % of Female Students | 100% |
| Admission Rate | 100% | Average Age of Entering College | 100% |
| % of Asian Students | 100% | | |

*Votes Column: The percentage of folds that selected the corresponding attributes.*

**Attribute Selection using Genetic Algorithm.** The Genetic Algorithm (GA) was the third option for attribute selection. The settings of the GA were as follows:

- Population size: 500
- Fitness function: Classification accuracy derived from Logistic Regression
- Selection Method: Tournament selection
- Crossover Type: Two-point crossover
- Crossover Rate: 0.6
- Mutation Rate: 0.03
- Stopping Criteria: 60 generations

In alignment with the prior two attribute selection approaches, 10-fold cross validation scheme in Weka was used. Attributes selected by at least six out of ten folds (60%) were selected, yielding 22 selected attributes in total (see Table 3).

**Table 3.** Selected Attributes Subset using Genetic Algorithm

| Attributes | Votes* | Attributes | Votes* |
|---|---|---|---|
| School Type | 60% | % of Asian Students | 100% |
| Predominant Degree Type | 70% | % of Hispanic Students | 100% |
| Student Size | 100% | % of Students whose Family Income classified Higher Middle | 80% |
| Instructional Expenditure per Student | 90% | % of Students whose Family Income classified Very High | 100% |
| Ratio between Part-time and Full-time Students | 100% | % of Students whose Parents have a Middle School Degree | 70% |
| Degree Completion Rate | 100% | % of Students whose Parents have a Post-High-School Degree | 60% |
| Admission Rate | 100% | % of Population from Students' Zip Codes over 25 with a Professional Degree | 100% |
| Average SAT Score | 90% | % of Female Students | 100% |
| Out-of-State Tuition | 100% | % of 1st Generation Students | 60% |
| % of White Students | 90% | Average Age of Entering College | 100% |
| % of Black Students | 60% | Average Debt | 70% |

*Votes Column: The percentage of folds that selected the corresponding attributes.*

**Comparisons among the three selected attribute subsets.** Logistic Regression and Support Vector Machine with Pearson VII function kernel were used to compare the performance of the three selected attribute subsets. Ten-fold cross validation was used to estimate the classification accuracy for each classification method. The individual classification results are presented in Table 4 and Table 5. Being the most selective attribute selection method (*9 attributes selected*), Forward Selection achieve acceptable F-measure. Although less selective (*22 attributes selected*), Genetic Algorithm outperformed the other two methods by both F-measure and accuracy.

**Table 4.** Comparisons among Three Selected Attribute Subsets Using Logistic Regression.

| Logistic Regression | Accuracy | Weighted Average | | |
|---|---|---|---|---|
| | | Precision | Recall | F-measure |
| Attribute Subset Selected by Filter Methods (N = 13) | 0.691 | 0.688 | 0.691 | 0.686 |
| Attribute Subset Selected by Forward Selection (N = 9) | 0.736 | 0.733 | 0.736 | 0.731 |
| Attribute Subset Selected by Genetic Algorithm (N = 22) | 0.746 | 0.746 | 0.746 | 0.745 |

**Table 5.** Comparisons among Three Selected Attribute Subsets Using Support Vector Machine with Pearson VII function kernel.

| Support Vector Machine with Pearson VII function kernel | Accuracy | Weighted Average | | |
|---|---|---|---|---|
| | | Precision | Recall | F-measure |
| Attribute Subset Selected by Filter Methods (N = 13) | 0.708 | 0.697 | 0.708 | 0.701 |
| Attribute Subset Selected by Forward Selection (N = 9) | 0.733 | 0.723 | 0.733 | 0.726 |
| Attribute Subset Selected by Genetic Algorithm (N = 22) | 0.755 | 0.745 | 0.755 | 0.747 |

## 3.2 Classification Using the Best Attribute Subset

Ten machine learning algorithms from five groups were applied to the best attribute subset, which is selected by the Genetic Algorithm (see Table 3). The five groups of algorithms include:
(1) Bayes-based algorithms [*Naive Bayes Update, Bayes Net*];
(2) Function-based algorithms [*Logistic Regression, Support Vector Machine (SVM) Multilayer Perceptron*];
(3) Instance-based algorithms [*Distance-weighted K-Nearest Neighbor (KNN)*];

(4) Tree-based algorithms [*J48 decision tree (J48), Multiclass Alternating Decision Tree (ADTree)*]; and
(5) Rule-based algorithms [*OneR, JRIP*].

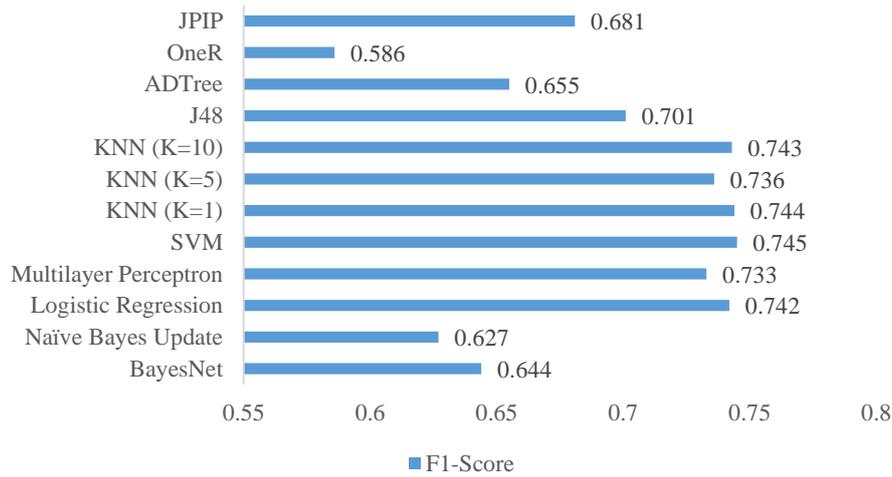

**Figure 1**. Performance Comparison among Single Learners Using the Best Attribute Subset.

The weighted average F1-score was chosen as the primary evaluation criterion, as there was an imbalance in the four income classes. A classifier that primarily guesses based on the majority class would achieve a small advantage in accuracy, but would perform worse in terms of F1-score. In addition, classification accuracy rate was used as the secondary evaluation criterion. Ten-fold cross validation was used for the estimate of both F1-score and accuracy rate. The results from each algorithm group are presented in Figure 1. The top three performers were identified as Support Vector Machine, and K-Nearest Neighbor with K equal to 1 and 10 respectively. The detailed classification results are presented in Table 6.

**Table 6.** Top Three Performers of Single Learners Using the Best Attribute Subset

| Algorithm | Accuracy | Weighted Average | | |
|---|---|---|---|---|
| | | Precision | Recall | F1-Score |
| Support Vector Machine (*kernel = Pearson VII function*) | 0.753 | 0.743 | 0.753 | 0.745 |
| K-Nearest Neighbor (*distance weight = 1/distance; K = 1*) | 0.745 | 0.744 | 0.745 | 0.744 |
| K-Nearest Neighbor (*distance weight = 1/distance; K = 10*) | 0.747 | 0.748 | 0.747 | 0.743 |

## 4 Discussion

Using the data from the College Scorecard [20], we selected the most important attributes predicting the 6-year post-graduation income of college students who used financial aid during their college time. More specifically, we compared three attribute selection methods, including filter methods, forward selection, and Genetic Algorithm, in terms of the classification accuracy on students' post-graduation income. In this process, we found that the attribute subset selected by the Genetic Algorithm outperformed the other two subsets when using logistic regression and support vector machine as the classification algorithm.

We wish to draw attention to attributes that were selected by at least two selection methods related to the socio-economic status of graduates (see Figure 2 and Figure 3). Higher numbers of students in a cohort who grew up in Zip code areas where over 25 percent of the population hold a Professional Degree was predictive of more college graduates likely being classified as High income. This trend is highly correlated with the relationship between the income of graduates and the ratio of high-income parents in the same cohort. The findings are in line with emerging evidence about the influence of geography or "where you grow up" on life outcomes. Chetty et al. [25] identified that areas with lower racial segregation and income inequality, but higher social capital[1] and family stability are associated with greater opportunities for intergenerational social mobility. Building on this work, our study suggests the importance geography for the post-graduation incomes in the case of neighborhood Professional Degree attainment.

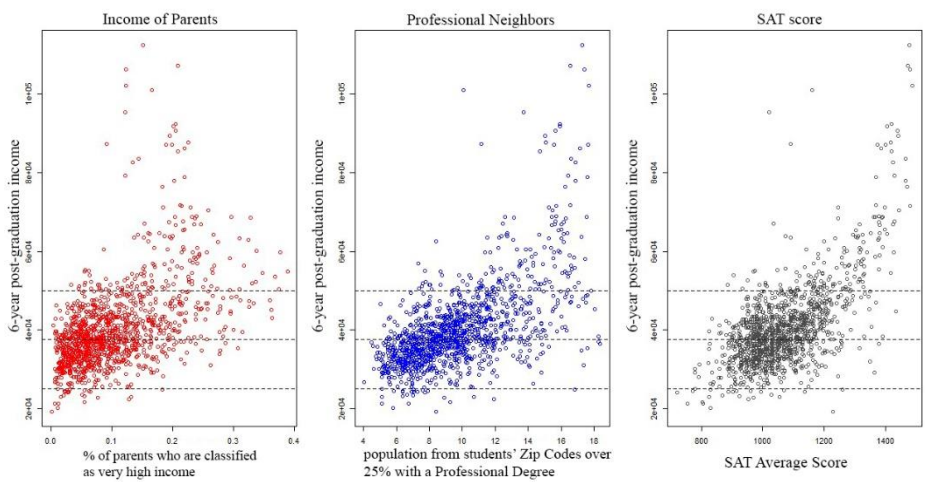

**Figure 2.** Relationships among six-year post-graduation income, percentage of parents who are classified as very high income, population from student's zip codes over 25% with a professional degree, and SAT score.

---

[1] Social capital represents trust, solidarity, and reciprocity in collective social interactions as well as participation in community organisations (Putnam, 2016).

The finding in terms of the relationship between SAT score and post-graduation income reinforces findings of prior research on post-graduation incomes [27]. It is important to note that all three relationships – Zip code areas where over 25% of the population hold a Professional Degree, parental income, and SAT scores – are highly correlated with each other. These findings suggest that neighbors who have attained a Professional Degree and parents with higher incomes may to be more able to provide an environmental that is conducive to educational development, which leads to more competitiveness in SAT. Also, students with higher SAT scores tend to attend more selective colleges and are therefore more likely to receive higher income after graduation (see Figure 2). This correlation echoes findings in the literature at a much larger scale using College Scorecard data. We would encourage researchers to conduct a mediational model to test relationships among neighborhood professional degree attainment, parents' income, SAT scores, and graduate incomes.

Moreover, as the percentage of students whose parents were a 1st generation college student increases, the post-graduation income of students is more likely to be classified as Low (see Figure 3). This finding may stem from families with a history of attending college being able to provide more informed educational and career related support to their children than their counterparts. Similarly, research has identified that first-generation college students are often handicapped in transition to the labor market by attending less selective institutions [4] and are more likely to encounter difficulties with their academic studies at college [28].

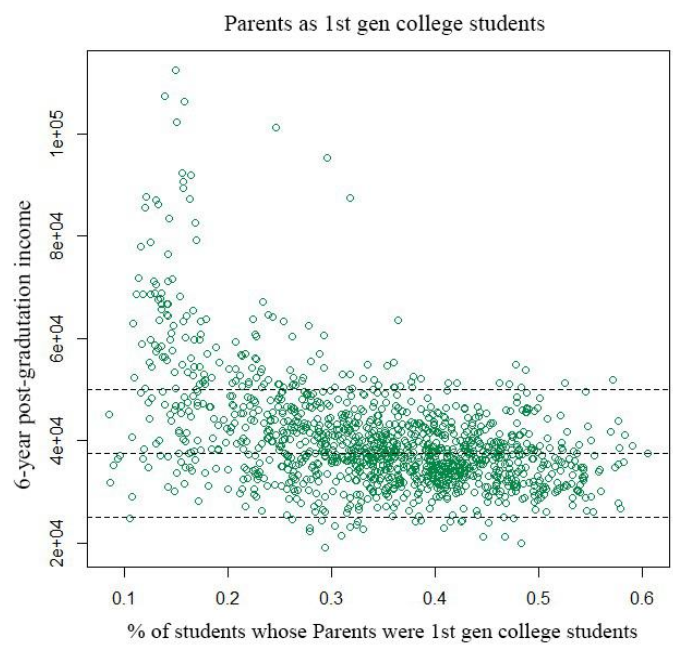

**Figure 3.** Relationship between six-year post-graduation income and percentage of students whose parents were 1<sup>st</sup> generation college graduates.

Our findings have important implications for social stratification research as they shed light on socio-economic inequalities in the graduate labor market. In other words, graduates from high socio-economic status backgrounds (i.e. neighborhoods with high levels of Professional Degree attainment, high parental income, and high family education) are more likely to receive higher post-graduation incomes. This is in accordance with the Effectively Maintained Inequality (EMI) model that predicts that as access to higher levels of education widens, students from higher socio-economic status groups will seek "horizontal differentiation" by accessing *qualitatively* superior or distinctive types of education that maintains their advantage in society [29, 30]. It is crucial to underline that we are *not* arguing that young people from low socio-economic status backgrounds should not attend higher education. Clearly, higher education remains an excellent "investment" for young people as they prepare for their career. Rather our findings illuminate social stratification among graduate populations in modern labor markets and question conventional wisdom of higher education being the "great equalizer" of life chances. We call for more research to further understand the dynamics behind such inequalities and to identify policy solutions to support college students from low socio-economic status backgrounds.

## 5    Concluding Remarks

A college degree is perhaps the single best means for an individual in U.S. to enhance their income in the labor market and achieve intergenerational social mobility [4]. It is clear, however, that a rising "graduate premium" has not meant that college graduates enjoy uniform access to gainful employment and high pay. Rather, the labor market for graduates is better viewed in terms of growing heterogeneity as the rewards have become unevenly distributed across the graduate population. This point is especially significant for low socio-economic status students given the steadily rising costs of college over the past decades and associated increases in student debt [7], and suggests a greater need for targeted support for low socio-economic status students *both* in accessing college and in transitions to the labor market. In this environment, the large scale of the College Scorecard dataset provides a highly important resource by enabling prospective college students (alongside their parents and school advisors) to access transparent and detailed information about labor market outcomes of prior cohorts. We hope that our research complements the dataset by offering more detailed insights into post-graduation incomes.

# References

1. Autor, D. H. Skills, Education, and the Rise of Earnings Inequality Among the 'Other 99 Percent. Science 344, 6186, 843-851 (2014).
2. Goldin, C, & Katz, L. F. The race between education and technology. Cambridge: Harvard University Press (2009)
3. Autor, D. H., Dorn, D., & Hanson, G. H. The China Shock: Learning from Labor-Market Adjustment to Large Changes in Trade. Annual Review of Economics, 8, 205-240 (2016).
4. Chetty, R., J. Friedman, E. Saez, N. Turner, & D. Yagan. Mobility report cards: The role of colleges in intergenerational mobility. Technical report: Stanford University (2017).
5. Hout, M. Social and economic returns to college education in the United States. Annual Review of Sociology, 38, 379-400 (2012).
6. National Center for Educational Statistics [NCES]. Percentage of 18- to 24-year-olds enrolled in degree-granting postsecondary institutions, by level of institution and sex and race/ethnicity of student: 1970 through 2015 Retrieved online from http://nces.ed.gov/programs/digest/d15/tables/dt15_302.60.asp?current=yes last accessed 2018/03/01.
7. Avery, C., & Turner, S. Student loans: Do college students borrow too much—or not enough?. The Journal of Economic Perspectives, 26(1), 165-192 (2012).
8. Wolff, E. N., Baumol, W. J., & Saini, A. N. A comparative analysis of education costs and outcomes: The United States vs. other OECD countries. Economics of Education Review, 39, 1-21 (2014).
9. Houle, J. N. Disparities in debt: Parents' socioeconomic resources and young adult student loan debt. Sociology of Education, 87(1), 53-69 (2014).
10. Brown, M., Haughwout, A., Lee, D., Scally, J., & Van Der Klaauw, W. Measuring student debt and its performance. Student loans and the dynamics of debt, 37-52 (2015).
11. Beaudry, P., Green, D. A., & Sand, B. M. The declining fortunes of the young since 2000. The American Economic Review, 104(5), 381-386 (2014).
12. Valletta, R. G. Recent Flattening in the Higher Education Wage Premium: Polarization, Skill Downgrading, or Both?. In Education, Skills, and Technical Change: Implications for Future US GDP Growth. University of Chicago Press (2017).
13. Federal Reserve Bank of New York. The Labor Market for Recent College Graduates. Retrieved online from https://www.newyorkfed.org/research/college-labor-market/index.html last accessed 2018/03/01.
14. Haughwout, A., Lee, D., Scally. J., & Van der Klaauw, W. Press Briefing on Student Loan Borrowing and Repayment Trends, 2015. Liberty Street Economics (2015). Retrieved online from http://www.montana.edu/cstoddard/documents/Loans_and_Performance.pdf last accessed 2018/03/01.
15. Hoekstra, M. The effect of attending the flagship state university on earnings: A discontinuity-based approach. The Review of Economics and Statistics, 91(4), 717-724 (2009)

16. Witteveen, D., & Attewell, P. The Earnings Payoff from Attending a Selective College. Social Science Research, (2017).
17. Dale, S., & Krueger, A. B. (Estimating the return to college selectivity over the career using administrative earnings data (No. w17159). National Bureau of Economic Research (2011).
18. Altonji, J. G., Arcidiacono, P., & Maurel, A. The analysis of field choice in college and graduate school: Determinants and wage effects (No. w21655). National Bureau of Economic Research (2015).
19. Kim, C., Tamborini, C. R., & Sakamoto, A. Field of study in college and lifetime earnings in the United States. Sociology of Education, 88(4), 320-339. (2015).
20. U.S. Department of Education. https://www.newyorkfed.org/research/college-labor-market/index.html last accessed 2018/03/01.
21. U.S. Census Bureau. Distribution of Personal Income 2010. Retrieved online from https://www.census.gov/2010census/data/  last accessed 2018/03/01
22. Beasley, J. E., & Chu, P. C. A genetic algorithm for the set covering problem. European Journal of Operational Research, 94(2), 392-404 (1996).
23. Goldberg, D. (1989). Genetic algorithms in optimization, search and machine learning. Reading: Addison-Wesley (1989).
24. Hall, M., Frank, E., Holmes, G., Pfahringer, B., Reutemann, P., & Witten, I. H. The WEKA data mining software: an update. ACM SIGKDD explorations newsletter, 11(1), 10-18 (2009).
25. Chetty, R., Hendren, N., Kline, P., & Saez, E. Where is the land of opportunity? The geography of intergenerational mobility in the United States. The Quarterly Journal of Economics, 129(4), 1553-1623 (2014).
26. Putnam, R. D. Our kids: The American dream in crisis. New York: Simon and Schuster (2016)
27. French, M. T., Homer, J. F., Popovici, I., & Robins, P. K. What you do in high school matters: High school GPA, educational attainment, and labor market earnings as a young adult. Eastern Economic Journal, 41(3), 370-386 (2015).
28. Pascarella, E. T., Pierson, C. T., Wolniak, G. C., & Terenzini, P. T. First-generation college students: Additional evidence on college experiences and outcomes. The Journal of Higher Education, 75(3), 249-284 (2004)
29. Lucas, S. R. Effectively maintained inequality: education transitions, track mobility, and social background effects. American Journal of Sociology, 106, 1642-1690 (2001).
30. Lucas S. R. & Byrne D. Effectively Maintained Inequality in Education: An Introduction. American Behavioral Scientist, 61(1), 3-7 (2017).

# Appendix A

The dataset analyzed in this study can be accessed at
https://collegescorecard.ed.gov/data/ .

30 potential attributes include:
*Group One: School information*
1. School Type (e.g. private school)
2. Predominant Awarded Degrees (e.g., bachelor degree)
3. Student Size
4. Instructional Expenditure per Student
5. Ratio between Part-time and Full-time Students
6. Degree Completion Rate
7. Average Faculty Salary

*Group Two: Admission information*
8. Admission Rate
9. Average SAT Score

*Group Three: Cost information*
10. In-State Tuition
11. Out-of-State Tuition

*Group Four: Student information*
12. Percentage of White Students
13. Percentage of Black Students
14. Percentage of Asian Students
15. Percentage of American Indian Students
16. Percentage of Hispanic Students
17. Percentage of Female Students
18. Percentage of First-Generation Students
19. Average Age of Entering College
20. Average Debt

*Group Five: Family and community information*
21. Percentage of Students whose Family Income was classified as Low
22. Percentage of Students whose Family Income was classified as Lower Middle
23. Percentage of Students whose Family Income was classified as Higher Middle
24. Percentage of Students whose Family Income was classified as High
25. Percentage of Students whose Family Income was classified as Very High
26. Percentage of Students whose Parents were 1st Generation College Student
27. Percentage of Students whose Parents Have a Middle School Degree
28. Percentage of Students whose Parents Have a High School Degree
29. Percentage of Students whose Parents Have a Post-High-School Degree
30. Population from Students' Zip Codes over 25% with a Professional Degree